\documentclass[12pt]{article}
\renewcommand{\theequation}{\thesection.\arabic{equation}}

\newlength{\extraspace}
\newlength{\extraspaces}
\newcommand{\be}{\begin{equation}
\addtolength{\abovedisplayskip}{\extraspaces}
\addtolength{\belowdisplayskip}{\extraspaces}
\addtolength{\abovedisplayshortskip}{\extraspace}
\addtolength{\belowdisplayshortskip}{\extraspace}}
\newcommand{\ee}{\end{equation}}

\newcommand{\ba}{\begin{eqnarray}
\addtolength{\abovedisplayskip}{\extraspaces}
\addtolength{\belowdisplayskip}{\extraspaces}
\addtolength{\abovedisplayshortskip}{\extraspace}
\addtolength{\belowdisplayshortskip}{\extraspace}}
\newcommand{\ea}{\end{eqnarray}}

\newcommand{\bd}{\begin{displaymath}
\addtolength{\abovedisplayskip}{\extraspaces}
\addtolength{\belowdisplayskip}{\extraspaces}
\addtolength{\abovedisplayshortskip}{\extraspace}
\addtolength{\belowdisplayshortskip}{\extraspace}}
\newcommand{\ed}{\end{displaymath}}

\newcounter{saveeqn}
\newcommand{\alpheqn}{\setcounter{saveeqn}{\value{equation}}%
 \stepcounter{saveeqn}\setcounter{equation}{0}%
 \renewcommand{\theequation}
     {\mbox{\thesection.\arabic{saveeqn}\alph{equation}}}}
\newcommand{\reseteqn}{\setcounter{equation}{\value{saveeqn}}%
  \renewcommand{\theequation}{\thesection.\arabic{equation}}}

\newcommand{\newsection}[1]{
\vspace{12mm}
\pagebreak[3]
\addtocounter{section}{1}
\setcounter{equation}{0}
\setcounter{subsection}{0}
\setcounter{footnote}{0}
\noindent{\bf \thesection. #1}
\nopagebreak
\medskip
\nopagebreak}

\newcommand{\newsubsection}[1]{
\vspace{0.8cm}
\pagebreak[3]
\addtocounter{subsection}{1}
\noindent{\it \thesubsection. #1}
\nopagebreak
\vspace{2mm}
\nopagebreak}

\newcommand{\eap}{e^{-2\alpha\phi}}
\newcommand{\half}{\frac{1}{2}}
\newcommand{\sd}{{\rm d}}
\newcommand{\ft}{\tilde{f}}
\newcommand{\rt}{\tilde{r}}

\newcommand{\at}{\tilde{\theta}}
\newcommand{\rr}{\rho\rho}
\newcommand{\te}{\theta =}
\newcommand{\fa}{\frac{2}{1+\alpha^2}}
\newcommand{\rp}{r_{+}}
\newcommand{\an}{\quad {\rm and} \quad}
\newcommand{\qc}{\quad , \quad}

\begin{document}
\addtolength{\baselineskip}{1.5mm}

\thispagestyle{empty}
\begin{flushright}
hep-th/0101221\\
\end{flushright}
\vbox{}
\vspace{2.5cm}

\begin{center}
{\LARGE{Black diholes with unbalanced magnetic charges
        }}\\[16mm]
{Y.C. Liang and Edward Teo}
\\[6mm]
{\it Department of Physics\\
National University of Singapore \\
Singapore 119260}\\[15mm]

\end{center}
\vspace{2cm}

\centerline{\bf Abstract}\bigskip \noindent
We present a technique
that can be used to generate a static, axisymmetric solution of
the Einstein-Maxwell-Dilaton equations from  a stationary,
axisymmetric solution of the vacuum Einstein equations. Starting
from the Kerr solution, Davidson and Gedalin have previously made
use of this technique to obtain a pair of oppositely charged,
extremal dilatonic black holes, known as a black dihole. In this
paper, we shall instead start from the Kerr-NUT solution. It will
be shown that the new solution can also be interpreted as a
dihole, but with the black holes carrying unbalanced magnetic
charges. The effect of the NUT-parameter is to introduce a net
magnetic charge into the system. Finally, we uplift our solution
to ten dimensions to describe a system consisting of $D6$ and
anti-$D6$-branes with unbalanced charges. The limit in which they
coincide agrees with a solution recently derived by Brax et al..

\newpage

\newsection{Introduction}

In 1983, Gross and Perry \cite{Gross} and Sorkin \cite{Sorkin}
derived  solutions of Kaluza-Klein theory  starting from existing
solutions to the four-dimensional vacuum Einstein equations. These
solutions were constructed by first analytically continuing the
seed solutions to the Euclidean  regime $t\to ix^5$, and adding on
an extra flat time direction. Solutions to the four-dimensional
Kaluza-Klein equations can then be obtained by compactifying the
five-dimensional spacetimes on $x^5$.

Using the self-dual Taub-NUT solution as seed, the authors
\cite{Gross, Sorkin} obtained a static solution  whose
Kaluza-Klein gauge field corresponds to that of a magnetic
monopole. This gauge field originates from the  $g_{t\varphi}$
term of the Taub-NUT solution,  with the NUT-parameter  attributed
the physical meaning of magnetic charge in the new solution.

Similarly, by using the Kerr solution as seed, a static  solution
whose gauge field describes that of a magnetic dipole was
constructed \cite{Gross}. In this case, the angular momentum  of
the Kerr solution is converted into a  parameter that
characterizes the dipole moment of the new solution. The latter is
the Kaluza-Klein analog  of the Bonnor solution \cite{Bonnor}
which describes a magnetic dipole in general relativity.

Now,  Kaluza-Klein theory is a special case of the more general
Einstein-Maxwell-Dilaton theory with arbitrary dilatonic coupling.
In the former, the coupling constant $\alpha$, defined in
(\ref{4daction}), is taken to be $\sqrt{3}$. However, in this
paper, we shall focus on the more general Einstein-Maxwell-Dilaton
theory where $\alpha$ can take on any real value.

In 1994, Davidson and Gedalin \cite{Davidson} generalized the
solution-generating technique to one that is valid for arbitrary
dilatonic coupling using a Ernst-type formalism. Starting from the
Kerr solution, they obtained the dilatonic generalization of the
Bonnor solution and interpreted the $\alpha=1$ case as one
exhibiting a two-dimensional black and white dihole structure.
However, Emparan \cite{Emparan} recently pointed out a flaw in
this interpretation and demonstrated that the solution (for
general $\alpha$) actually describes  a static pair of oppositely
charged, extremal dilatonic black holes, which he calls a black
dihole.

We will start off in Sec.~2 by presenting the solution-generating
technique that was first used by Davidson and Gedalin to obtain
the dilatonic generalization of the Bonnor solution. Since this
technique was only briefly described in their original article
\cite{Davidson}, we will provide details of it for completeness.
To the best of our knowledge, these details cannot be found
elsewhere in the literature.

In Sec.~3, we will make use of this technique to derive our dihole
solution starting from the Kerr-NUT solution \cite{Demianski}; the
latter is a generalization of the Kerr solution which includes an
additional NUT-parameter. As such, our solution contains all the
above-mentioned ones as special cases.

This will be followed by a standard analysis, that was first used
by Emparan \cite{Emparan}, to investigate the physical nature of
the new  solution. We will demonstrate that it describes a pair of
extremal dilatonic black holes carrying unbalanced magnetic
charges, i.e.~each of the black holes carries a magnetic charge
that is of different sign as well as magnitude. In contrast to a
dipole, this system will have a non-trivial net magnetic charge.
We will continue to refer to this solution as a dihole.

In Sec.~4, we will discuss the embedding of our solution in Type
IIA superstring theory, in which it would describe a static
configuration of $D6$ and anti-$D6$-branes with unbalanced
charges. We will then show that the coincident limit agrees with a
solution recently presented by Brax et al.~\cite{Brax}.

\newsection{Solution-generating technique}

The idea of generating a solution of the Einstein-Maxwell
equations starting from a solution of the vacuum Einstein
equations is not a new one (see for example Kramer et
al.~\cite{Kramer} or Islam~\cite{Islam}). In this section, we will
focus on a different technique which generates a static,
axisymmetric solution of the Einstein-Maxwell-Dilaton equations
starting from a stationary, axisymmetric solution of the vacuum
Einstein equations.

Einstein-Maxwell-Dilaton theory in four dimensions has the following action integral:
\be
\label{4daction}
\int \sd^4x\sqrt{-g}\left(R-2(\nabla\phi)^2-\eap F^2\right),
\ee
where $R$ is the Ricci scalar, $\phi$  the dilaton field and $F_{ab}$  the electromagnetic field tensor. For the special case $\alpha=0$, we recover Einstein-Maxwell theory; for $\alpha=1$, the action describes the low energy dynamics of string theory; and for $\alpha=\sqrt{3}$, we have  Kaluza-Klein theory as mentioned in the introduction. However, we shall keep $\alpha$ general in what follows. By varying this action with respect to the metric,  gauge field and  dilaton field, we obtain the respective field equations:
\alpheqn
\ba
\label{eqnmnR}&&R_{ab}= 2\nabla_a\phi\nabla_b\phi + \eap\left(2 F_{ac}{F_b}^c-\half g_{ab} F^2\right),\\
\label{eqnmnA}&&\nabla_a\,(e^{-2\alpha\phi}F^{ab})= 0,\\
\label{eqnmnP}&&\nabla^2\phi+\frac{\alpha}{2}\eap F^2 =0.
\ea
\reseteqn

Now, we are looking for a static, axisymmetric solution to the field equations. Recall that any such spacetime can be cast in the Weyl-Papapetrou form \cite{Islam}:
\ba
\label{metric_ansatz} \sd s^2 &=& -f\sd t^2 + l\sd\varphi^2 +e^\mu (\sd\rho^2+\sd z^2).
\ea
Furthermore, we choose a purely magnetic ans\"atz $A \equiv A_\varphi$ with  all other components of $A_a$ vanishing. It is also understood that $f$, $l$, $\mu$, $A$ and $\phi$ are functions of $\rho$ and $z$ only.

With this ans\"atz, we can now evaluate the left- and right-hand
sides of (\ref{eqnmnR}). Defining $D^2\equiv fl $, and by
considering the combination $e^\mu
D^{-1}(lR_{tt}-fR_{\varphi\varphi})$, we obtain \bd
\frac{\partial^2 D}{\partial\rho^2} +\frac{\partial^2 D}{\partial
z^2}=0. \ed A simple solution to this equation is given by
$D=\rho$ \cite{Islam}. With this choice, we may evaluate the
$R_{tt}$ equation to arrive at \be
\label{f}f(f_{\rho\rho}+f_{zz}+\rho^{-1}f_\rho)-{f_\rho}^2-{f_z}^2
= 2\rho^{-2}f^3\eap ({A_\rho}^2+{A_z}^2), \ee where subscripts
$\rho$ and $z$ indicate partial derivatives of the corresponding
function with respect to these variables. Similarly, by evaluating
the Ricci components $R_{\rho\rho}-R_{zz}$ and $R_{\rho z}$, we
obtain respectively \alpheqn \ba
\label{mur}\mu_\rho&=&-f^{-1} f_\rho +\half\rho f^{-2}({f_\rho}^2-{f_z}^2) + 2\rho({\phi_\rho}^2-{\phi_z}^2)+2\eap\rho^{-1}f({A_\rho}^2-{A_z}^2),\\
\label{muz}\mu_z&=&- f^{-1} f_z + \rho f^{-2} f_\rho f_z + 4\rho\,\phi_\rho\phi_z + 4\eap\rho^{-1} f A_\rho A_z.
\ea
\reseteqn

Finally, we obtain from (\ref{eqnmnA}) and (\ref{eqnmnP}) the gauge field and dilaton equations respectively:
\ba
\label{gauge_eqn} &&A_{\rho\rho} + A_{zz} - \rho^{-1}A_\rho = 2\alpha\left( A_\rho \phi_\rho + A_z\phi_z\right) -f^{-1}\left( A_\rho f_\rho + A_z f_z\right),\\
\label{dilaton_eqn} &&\phi_{\rho\rho} + \phi_{zz} + \rho^{-1}\phi_\rho = -\alpha\rho^{-2} f\eap\left({A_\rho}^2+{A_z}^2\right).
\ea

The crucial step now is to realize that if we set
\be
\label{correspondence}\ft^2=f\eap\qquad {\rm and}\qquad w=i\sqrt{1+\alpha^2}A,
\ee
then (\ref{f}) and (\ref{gauge_eqn}) respectively become
\ba
\label{ff}&\ft (\ft_{\rr}+\ft_{zz}+\rho^{-1}\ft_\rho)-{\ft_\rho}\,^2-{\ft_z}\,^2 +\rho^{-2}\ft^4({w_\rho}^2+{w_z}^2)=0,\\
\label{fgauge}&\ft(w_{\rr}+w_{zz}-\rho^{-1}w_\rho)+2(w_\rho\ft_\rho+w_z\ft_z)=0.
\ea
These equations are precisely the same as those derived from the vacuum Einstein equations for a stationary, axisymmetric metric
\be
\label{metric_rotating} \sd s^2= -\ft(\sd t - w \sd\varphi)^2 + \rho^2 \ft^{-1}\sd\varphi^2 + e^{\tilde{\mu}}(\sd\rho^2+\sd z^2),
\ee
using a Ernst-type formalism (c.f.~Eqn.~(2.12a) and (2.12b) of Islam~\cite{Islam}).
For every such solution to the vacuum Einstein equations, we can therefore find a corresponding solution to the Einstein-Maxwell-Dilaton equations via (\ref{correspondence}). Nevertheless, one should be aware that this procedure would in general generate an imaginary gauge field. Thus, a real solution to the Einstein-Maxwell-Dilaton equations can be generated via this method only if an analytic continuation of the parameter(s) in the seed solution is possible.

Finally, we see that the dilaton equation (\ref{dilaton_eqn}) can be written as
\be
\label{fdilaton}\phi_{\rho\rho} + \phi_{zz} + \rho^{-1}\phi_\rho = -\alpha\rho^{-2}\ft^2 \left({A_\rho}^2+{A_z}^2\right),
\ee
which, using (\ref{ff}), admits the solution
\be\label{dilaton_expression}
\phi=-\frac{\alpha}{1+\alpha^2}\ln\ft,
\ee
up to the addition of a harmonic function $\tilde{\phi}$ satisfying $\tilde{\phi}_{\rho\rho} + \tilde{\phi}_{zz} + \rho^{-1}\tilde{\phi}_\rho =0$.
For the choice $\tilde{\phi}=0$, we have
\be\label{f_expression}
f=\ft^\fa.
\ee
These expressions, together with that for the gauge field obtained from ({\ref{correspondence}), can then be used to deduce $\mu$ via (\ref{mur}) and (\ref{muz}). This completes our derivation of the static, axisymmetric solution to the field equations (\ref{eqnmnR})\,--\,(\ref{eqnmnP}).

We remark that this technique is just a dilatonic generalization
of Theorem~30.8 in Kramer et al.~\cite{Kramer}, which Bonnor
\cite{Bonnor} used to generate (from the Kerr solution) the
well-known magnetic dipole solution. Using the same solution as
seed, Davidson and Gedalin \cite{Davidson} have made use of the
above technique to generate the dilatonic generalization of the
Bonnor solution. If we begin with the self-dual Taub-NUT solution
instead, we would obtain a generalization of the
Gross-Perry-Sorkin \cite{Gross, Sorkin} monopole solution to
arbitrary dilatonic coupling, which can also be interpreted as an
extremal dilatonic black hole~\cite{Gibbons}.

\newsection{Dihole solution with unbalanced charges}

\newsubsection{Derivation of solution}

In this paper, we will start from the Kerr-NUT (or Demia\'nski-Newman) solution \cite{Demianski}:
\ba\label{Demianski}
\sd s^2&=&-\bar{\Lambda}\left(\sd t+2\frac{a\sin^2\theta\,(mr+l^2)+l \bar{\Delta}\cos\theta}{\bar{\Delta}-a^2\sin^2\theta}\sd\varphi \right)^2 \nonumber\\
\label{metric_DM}&&+\bar{\Lambda}^{-1}\left[ (\bar{\Delta}-a^2\sin^2\theta)\left( \frac{\sd r^2}{\bar{\Delta}}+ \sd\theta^2\right)+\bar{\Delta}\sin^2\theta\,\sd\varphi^2 \right],
\ea
where
\bd
\bar{\Lambda}\equiv\frac{\bar{\Delta} - a^2\sin^2\theta}{r^2+(a\cos\theta+l)^2}\quad {\rm and}\quad \bar{\Delta}\equiv r^2-2mr+a^2-l^2.
\ed
In these expressions, $a$ is the angular momentum and $l$ is the NUT-parameter. Note that this solution contains both the Kerr and Taub-NUT solutions as  special cases with $l=0$ and $a=0$ respectively.

By comparing the line element (\ref{metric_DM}) with (\ref{metric_rotating}), we obtain
\be
\ft = \bar{\Lambda}\an w = -2\frac{a\sin^2\theta\,(mr+l^2)+l\bar{\Delta}\cos\theta}{\bar{\Delta}-a^2\sin^2\theta}.
\ee
To ensure that the resulting gauge field is real, we perform the analytic continuation $a\to ia$ and $l\to il$. Using (\ref{correspondence}), (\ref{dilaton_expression}) and (\ref{f_expression}), we obtain
\ba
\label{gauge}A_\varphi&=&-\frac{2}{\sqrt{1+\alpha^2}}\frac{a\sin^2\theta\,(mr-l^2)+l\Delta\cos\theta}{\Delta+a^2\sin^2\theta},\\
\label{dilaton}\phi&=&-\frac{\alpha}{1+\alpha^2}\ln\left[ \frac{\Delta+a^2\sin^2\theta}{\Sigma}\right],\\
\label{fff}f&=&\left[ \frac{\Delta+a^2\sin^2\theta}{\Sigma}\right]^\fa,
\ea
where
\be\label{DeltaSigma}\Delta\equiv r^2-2mr-a^2+l^2\qquad\an\qquad \Sigma\equiv r^2-(a\cos\theta+l)^2.\ee

Substituting (\ref{gauge})\,--\,(\ref{fff}) into (\ref{mur}) and (\ref{muz}), and by transforming from the Boyer-Lindquist-type coordinates ($r$, $\theta$) to  cylindrical coordinates ($\rho$, $z$) via  \cite{Carmeli}
\bd
\rho=\sqrt{r^2-2mr-a^2+l^2}\,\sin\theta ,\qquad z=(r-m)\cos\theta,
\ed
we obtain by quadrature\footnote{For practical reasons, the integration was actually performed after further transforming to  prolate spheroidal coordinates \cite{Carmeli}:
$$ x=\frac{r-m}{\sqrt{m^2+a^2-l^2}}\qquad\an\qquad y=\cos\theta. $$
} the expression for $\mu$. We finally arrive at the new solution
\be\label{newmetric}
\sd s^2=\Lambda^{\fa}\left\{-\sd t^2+\frac{\Sigma^{\frac{4}{1+\alpha^2}}}{\left[\Delta + (m^2+a^2-l^2)\sin^2\theta\right]^{\frac{3-\alpha^2}{1+\alpha^2}}}\left(\frac{\sd r^2}{\Delta}+\sd\theta^2\right)\right\}+\frac{\Delta\sin^2\theta}{\Lambda^{\fa}}\sd\varphi^2,
\ee
where
\bd
\Lambda\equiv\frac{\Delta + a^2\sin^2\theta}{\Sigma},
\ed
with $A_\varphi$ and $\phi$ given in (\ref{gauge}) and (\ref{dilaton}) respectively.

\newsubsection{Physical properties of the solution}

Recently, the special case ($l=0$) of the above solution was
analyzed in detail by Emparan \cite{Emparan}. In this section, we
will perform a similar analysis on our solution and  arrive at the
conclusion that it describes a pair of extremal dilatonic black
holes with unbalanced charges lying on the symmetry axis.

We begin by  highlighting that, in addition to being static and
axisymmetric, the solution is asymptotically flat. This is in
contrast to our seed solution (\ref{Demianski}) whose
$g_{t\varphi}$ term does not vanish in the asymptotic limit.  This
unphysical nature of the Kerr-NUT solution has thus been removed
in the new solution.

A study of the asymptotic behavior of $g_{tt}$ and $A_\varphi$
also reveals that the total mass of the solution is
$M=\frac{2m}{1+\alpha^2}$ whereas  the net magnetic charge of the
solution is $\bar{Q}=\frac{2l}{\sqrt{1+\alpha^2}}$. Thus the
NUT-parameter $l$ governs the monopole field strength of the
solution at far field. Without loss of generality, we shall
restrict ourselves to non-negative $l$ corresponding to
non-negative net magnetic charge.

We shall now examine the singularities of the metric. By
evaluating the curvature invariant $R_{abcd}R^{abcd}$, it can  be
checked that for $l\le m$\footnote{This range for $l$ will be
justified below.}, the ``outermost'' curvature singularities are
located at the two points: \be\label{singularity} r=\rp\equiv
m+\sqrt{m^2+a^2-l^2}\qc\te0,\,\pi. \ee We can then follow a
similar analysis as in \cite{Emparan} to show that the axis of
symmetry consists of the three line segments $\te 0$, $r=\rp$ and
$\te \pi$, and the singularities given by (\ref{singularity}) are
merely the joints between these segments (c.f.~Fig~1 of
\cite{Emparan}).

In order to better understand the nature of these singularities,
we first note that the proper distance between the two
singularities increases as $2a$ when $a\to\infty$. It can also be
shown that (for $\alpha\ne 0$) the proper distance vanishes when
$a\to 0$.\footnote{For $\alpha=0$ however, the proper distance
remains infinite in this limit. It would be clear later that this
is due to the well-known fact that extremal Reissner-Nordstr\"om
black holes have throats of infinite length.} Thus, the parameter
$a$ serves as a measure of the distance between the two
singularities.

Bearing these facts in mind, we may now further investigate the
two singularities by adopting the following transformation
\cite{Sen,Emparan}:
\be\label{transformation} r=\rp +
\frac{\rt}{2}(1+\cos\at)\an\sin^2\theta=\frac{\rt(1-\cos\at)}{\sqrt{m^2+a^2-l^2}},
\ee
on the metric (\ref{newmetric}), while taking the limit
$a\to\infty$. Physically, this is tantamount to pushing one of the
singularities to a large distance and studying the geometry of the
remaining singularity. After carrying out the transformation, we
obtain
\alpheqn\ba
\label{zoom_out_metric}\sd s^2&\to& -\left(1+\frac{|Q|}{\rt}\right)^{-\fa}\sd t^2+\left(1+\frac{|Q|}{\rt}\right)^\fa\left[\sd\rt^2+\rt^2(\sd\at^2+\sin^2\at \,\sd\varphi^2)\right],\\
\label{zoom_out_gauge}A_\varphi&\to&\frac{Q\cos\at}{\sqrt{1+\alpha^2}},\\
\label{zoom_out_phi}\phi&\to&-\frac{\alpha}{1+\alpha^2}\ln\left(1+\frac{|Q|}{\rt}\right),
\ea where \be\label{Q} Q|_{\theta=0}=m -l\an Q|_{\theta=\pi}=-m-l.
\ee \reseteqn
This limiting form is just that of an extremal
dilatonic black hole, with the (singular) horizon located at
$\rt=0$ ($r=\rp$)~\cite{Gibbons}.

We could also perform the transformation (\ref{transformation}) on
the metric (\ref{newmetric}) without taking the limit of large
$a$. For small $\rt$, it enables us to investigate the geometry
near to the two singularities. In this limit, the geometry reduces
to the near-horizon limit of an extremal dilatonic black hole.
However, the horizon will  no longer be spherically symmetric due
to the presence of the other black hole. One can readily calculate
the relevant distortion factors following~\cite{Emparan}.

It can therefore be seen that at the ends of the segment $r=\rp$,
there lie two extremal dilatonic black holes  carrying unbalanced
magnetic charges. With the aid of Gauss's law, we may also deduce
from (\ref{zoom_out_gauge}) that the black hole at ($r$,
$\theta$)=($\rp$, 0) and ($\rp$, $\pi$) carries a magnetic charge
of $\frac{l-m}{\sqrt{1+\alpha^2}}$ and
$\frac{l+m}{\sqrt{1+\alpha^2}}$ respectively. As expected, the sum
of these magnetic charges matches exactly with the net charge of
the solution obtained above.

The next step is to determine if there are any conical
singularities along the different segments of the symmetry axis.
Assuming that the coordinate $\varphi$ has its usual periodicity
along the symmetry axes $\te0$ or $\pi$, it can be checked that
the conical excess along $r=\rp$ is given by
\be\label{conical_singularities}
\delta_{(\rp)}=2\pi\left[\left(1+\frac{m^2-l^2}{a^2}\right)^\fa-1\right].
\ee
As was pointed out in \cite{Emparan}, this conical excess can
be understood physically as the presence of a strut along $r=\rp$,
which provides the necessary internal stress to counterbalance the
attraction between the unbalanced-charged black holes.

To an observer located at $r>\rp$, the only observable physical
entities are thus the two black holes located at the  ends of the
segment $r=\rp$; when $0<\theta<\pi$, the region $r<\rp$ is
inaccessible due to the presence of conical singularities; when
$\te 0$ or $\pi$, all other singularities are located at $r<\rp$,
i.e.~enclosed within the horizon.

{} Now, note from (\ref{Q}) and (\ref{conical_singularities}) that
the magnetic charge of the black hole at ($\rp$, 0), as well as
the conical singularity along the segment $r=\rp$, vanishes when
$l=m$. To understand the physical nature of this special case, we
first note from (\ref{zoom_out_metric}) that  the masses of the
black holes are $ m|_{\theta=0}=  \frac{m-l}{1+\alpha^2}$ and
$m|_{\theta=\pi} = \frac{m+l}{1+\alpha^2}$ respectively. When
$l=m$, the mass of the black hole at $\theta=0$ vanishes whereas
that of the one at $\theta=\pi$ becomes the total mass of the
solution.  Intuitively, we can thus think of the increase of $l$
(from zero) as a physical process whereby the mass\footnote{In the
process of increasing the value of $l$, charge is being
transferred as well since the black holes are extremal.} of the
first black hole is transferred adiabatically to the second.

Indeed, this can seen by performing the  following transformation
on the line element (\ref{newmetric}) when $l=m$:} \bd \tilde{r} =
r-m+a\cos\theta \an \sin^2\tilde{\theta} =
\frac{(r-m)^2-a^2}{(r-m+a\cos\theta)^2}\sin^2\theta. \ed The
resulting line element is \bd \sd
s^2=-\left(1+\frac{2m}{\rt}\right)^{-\fa}\sd
t^2+\left(1+\frac{2m}{\rt}\right)^\fa[\sd\rt^2+\rt^2(\sd\tilde{\theta}^2
+ \sin^2\tilde{\theta}\,\sd\varphi^2)], \ed
which clearly
describes the geometry of an extremal dilatonic black hole with
mass $2m$ (c.f.~(\ref{zoom_out_metric})). Note that although the
transformation depends on $a$, the resulting line element does
not; this is expected because $a$ no longer carries any physical
meaning when there is only one black hole left in the
system.\footnote{The $a=0$ case, corresponding to the self-dual
Taub-NUT solution, was precisely what Gross and Perry \cite{Gross}
and Sorkin \cite{Sorkin} considered to obtain their Kaluza-Klein
monopole solution.} Notice that if we attempt to increase $l$
beyond $m$, the black hole located at ($\rp$, 0) would attain a
negative mass and thus become a naked singularity; at the other
end of the segment $r=\rp$, the mass of the black hole would
exceed the total mass of the solution. This is clearly an
unphysical situation; thus we shall restrict ourselves to values
of $l$ that are less than or equal to $m$.

When  $l=0$, it is well known \cite{Emparan} that the conical
singularity along $r=\rp$ can also be removed by introducing an
external magnetic field tuned to the appropriate strength; this
was achieved by performing a dilatonic generalization of the
Harrison transformation \cite{Dowker:1993} on the $l=0$ case of
our solution (\ref{newmetric}), (\ref{gauge}) and (\ref{dilaton}).
For general $l$, the transformation yields
\alpheqn\begin{eqnarray} \sd s^2&=&\Lambda'^{\fa} \left\{-\sd t^2
+\frac{\Sigma^{\frac{4}{1+\alpha^2}}}{\left[\Delta +
(m^2+a^2-l^2)\sin^2\theta\right]^{\frac{3-\alpha^2}{1+\alpha^2}}}
\left(\frac{\sd r^2}{\Delta}+\sd\theta^2\right)
\right\}+\frac{\Delta\sin^2\theta}{\Lambda'^{\fa}}\sd\varphi^2,\nonumber\\
\label{harrison_metric}\\
\label{harrison_gauge}A_\varphi&=&-\frac{1}{\Sigma\Lambda'}\biggl\{\frac{2}{\sqrt{1+\alpha^2}}[a\sin^2\theta\,(mr-l^2)+l\Delta\cos\theta] \nonumber\\
&&\hspace{1.41cm}-\frac{B}{2}[\sin^2\theta\,(r^2-a^2-l^2)^2+\Delta(a\sin^2\theta-2l\cos\theta)^2]\biggr\},\\
\label{harrison_dilaton}\phi&=&-\frac{\alpha}{1+\alpha^2}\ln\Lambda',
\end{eqnarray} \reseteqn
where $\Delta$ and $\Sigma$ are the same as above,
\ba
\Lambda'&\equiv&\frac{1}{\Sigma}\biggl\{\Delta+a^2\sin^2\theta-2B\sqrt{1+\alpha^2}[a\sin^2\theta\,(mr-l^2)+l\Delta\cos\theta] \nonumber\\
\label{Lambda'}&&\hspace{0.65cm}+\frac{1}{4}B^2(1+\alpha^2)[\sin^2\theta\,(r^2-a^2-l^2)^2+\Delta(a\sin^2\theta -2l\cos\theta)^2]\biggr\},
\ea
and $B$ is a new parameter governing the strength of the external magnetic field.

The values of $B$ that would remove the conical singularity along
$r=\rp$ are now \bd
B_\pm=\frac{1}{\sqrt{1+\alpha^2}}\frac{a\pm\sqrt{m^2+a^2-l^2}}{m\rp-l^2}.
\ed Of these two possible values, $B_+$ is unphysical as it
remains non-zero in the limit of large $a$ \cite{Emparan}.
Therefore, the only physically sensible $B$ that would remove the
conical singularity along $r=\rp$ is given by $B_-$.

However, in contrast to the $l=0$ case \cite{Emparan}, the conical
singularities at the other axes of symmetry no longer vanish for
this choice of $B$. If we assume that the coordinate $\varphi$ has
its usual periodicity along the symmetry axis $r=\rp$, it can be
shown that along the $\theta=0$ segment, there is a conical {\em
deficit} of
\begin{displaymath}
\delta_{(0)}=2\pi\left\{1-\left[1-l\left(\frac{a-\sqrt{m^2+a^2-l^2}}{m\rp-l^2}\right)
\right]^{-\frac{4}{1+\alpha^2}}\right\},
\end{displaymath}
corresponding to a cosmic string; whereas along the $\theta=\pi$
segment, there is a conical {\em excess} of
\begin{displaymath}
\delta_{(\pi)}=2\pi\left\{\left[1+l\left(\frac{a-\sqrt{m^2+a^2-l^2}}{m\rp-l^2}\right)
\right]^{-\frac{4}{1+\alpha^2}}-1\right\},
\end{displaymath}
corresponding to a strut. Hence, for an unbalanced dihole, it is
impossible to remove the conical singularities along the segments
$\theta=0$ and $\theta=\pi$ simultaneously with that along
$r=\rp$, by tuning the strength of the external magnetic field.
Physically, this is expected due to the asymmetry in the
distribution of charges among the two black holes.

Finally, we shall remark that instead of a magnetic dihole
solution, an electric dihole solution can be obtained by dualizing
the magnetic field strength tensor ${F}_{ab}$ via
\be\label{duality_transformation} \phi'=-\phi,\qquad
{F'}_{ab}=\frac{e^{-2\alpha\phi}}{2}\epsilon_{abcd}F\,^{cd}. \ee
Applying this transformation to the Harrison-transformed solution
(\ref{harrison_metric})\,--\,(\ref{harrison_dilaton}), we obtain a
solution which describes an electrically charged dihole immersed
in an external  electric field. The corresponding gauge field is
given by \ba\label{electricgauge}
A'\,_t&=&(r-3m)B\cos\theta -\frac{\sqrt{1+\alpha^2}}{2}B^2\left[ ma\cos\theta\,(2+\sin^2\theta)+l(r-3m)(1+\cos^2\theta)\right]\nonumber\\
&&+\frac{2}{\sqrt{1+\alpha^2}}\frac{ma\cos\theta-l(r-m)}{\Sigma}\left[1+\frac{\sqrt{1+\alpha^2}}{2}B(a\sin^2\theta-2l\cos\theta)\right]^2,
\ea
with all other components vanishing. In the special case $l=0$, the above expression reduces to  that  given by Chattaraputi et al.~\cite{Chattaraputi}. Note that $B$ now governs the strength of the external electric field. When $B=0$, (\ref{electricgauge}) reduces to
\be
A'\,_t=\frac{2}{\sqrt{1+\alpha^2}}\frac{ma\cos\theta-l(r-m)}{\Sigma},
\ee
which clearly asymptotes to the gauge field generated by an electric point source, with charge  $\frac{-2l}{\sqrt{1+\alpha^2}}$,  located at the origin.

\newsection{$D6$-anti-$D6$-brane configuration}

Now, for certain values of $\alpha$, the action (\ref{4daction})
emerges from string theory when compactified down to four dimensions.
In such cases the four-dimensional dihole solutions can be reinterpreted
in terms of brane-anti-brane configurations in ten dimensions. Perhaps the most important example is when the Kaluza-Klein dipole is uplifted to ten dimensions, to describe a $D6$-anti-${D6}$-brane configuration in Type IIA
superstring theory \cite{Sen}.

In the string frame, the solution describing a pair of $D6$-branes with opposite but unbalanced magnetic charges, immersed in a non-trivial magnetic field, is given by (in standard string theory conventions \cite{Brax})
\alpheqn
\ba
\label{D6D6}
\sd s^2&=&\Lambda'\,^{1\over2}\left\{-\sd t^2+\sd x_1^2+ \cdots+\sd x_6^2+\Sigma\left(\frac{\sd r^2}{\Delta}+\sd\theta^2\right)\right\}+\frac{\Delta\sin^2\theta}{\Lambda'\,^{1\over2}}\sd\varphi^2,\\
A_\varphi&=&-\frac{2}{\Sigma\Lambda'}\biggl\{a\sin^2\theta\,(mr-l^2)+l\Delta\cos\theta\nonumber\\
&&\hspace{1.3cm}-\frac{B}{2}\left[\sin^2\theta\,(r^2-a^2-l^2)^2+\Delta(a\sin^2\theta-2l\cos\theta)^2\right]\biggr\},\\
\label{string_dilaton}\phi&=&-{3\over4}\ln\Lambda',
\ea
\reseteqn
where we now have,
\begin{eqnarray*}
\Lambda'&=&\frac{1}{\Sigma}\biggl\{\Delta+a^2\sin^2\theta-4B[a\sin^2\theta\,(mr-l^2)+l\Delta\cos\theta] \\
&&\hspace{0.65cm}+B^2[\sin^2\theta\,(r^2-a^2-l^2)^2+\Delta(a\sin^2\theta -2l\cos\theta)^2]\biggr\},
\end{eqnarray*}
with $\Delta$ and $\Sigma$ given in (\ref{DeltaSigma}).  The geometry of the individual $D6$-branes, located at $(r,\theta)=(r_+,0)$ and $(r_+,\pi)$, can be recovered by performing the coordinate transformation (\ref{transformation}) on the above solution. This solution contains, as a special case, the solution considered by Sen in~\cite{Sen}.

As in the four-dimensional situation, if the external magnetic field is switched off by setting $B=0$, the branes coincide when $a=0$. In this
limit, the solution simplifies to
\alpheqn
\ba
\label{coincident_metric}
\sd s^2&=&\left(\frac{r^2-2mr+l^2}{r^2-l^2}\right)^{\half}(-\sd t^2+\sd x_1^2+ \cdots+\sd x_6^2) \cr
&&+ \left(\frac{r^2-l^2}{r^2-2mr+l^2}\right)^{\half}[\sd r^2 + (r^2-2mr+l^2)(\sd\theta^2+\sin^2\theta\,\sd\varphi^2)],\\
\label{coincident_gauge}A_\varphi&=&-2l\cos\theta\,,\\
\label{coincident_dilaton}\phi&=&-{3\over4}\ln\left(\frac{r^2-2mr+l^2}{r^2-l^2}\right),
\ea
\reseteqn
which describes a spherically symmetric\footnote{It is curious to note
that the coincident limit of the four-dimensional dihole (\ref{newmetric}) solution is spherically symmetric only
when $\alpha=\sqrt{3}$.} six-brane source carrying a monopole charge $l$. Recently, Brax et al.~\cite{Brax} presented a supergravity solution that corresponds to $N$ $Dp$-branes coinciding with $\overline{N}$ anti-${Dp}$-branes, with $N\neq\overline{N}$ in general. We will now establish a correspondence between our solution and theirs when $p=6$.

The solution of \cite{Brax}, after transforming to the string frame, is given by
\alpheqn
\ba
\label{Brax_metric}\sd s^2&=&e^{\frac{\phi}{2}}\left\{e^{2A(\rt)}\,(-\sd t^2+\sd x_1^2+ \cdots+\sd x_6^2)+e^{2B(\rt)}\,[\sd \rt^2+ \rt^2(\sd\theta^2+\sin^2\theta\,\sd\varphi^2)]\right\}, \qquad\\
\label{Brax_gauge}A_{t1\ldots 6}&=&\sqrt{c_2^2-1}\frac{\sinh k h(\rt)}{\cosh k h(\rt)-c_2\sinh k h(\rt)},\\
\label{Brax_dilaton}\phi&=& \frac{7}{16}c_1 h(\rt) -\frac{3}{4}\ln [\cosh k h(\rt)-c_2\sinh k h(\rt)],
\ea
\reseteqn
where
\begin{eqnarray*}
A(\rt)&\equiv&-\frac{3}{64}c_1 h(\rt) -\frac{1}{16}\ln [\cosh k h(\rt)-c_2\sinh k h(\rt)],
\\
B(\rt)&\equiv&\ln\left(1-\frac{r_0^2}{\rt^2}\right)+\frac{21}{64}c_1h(\rt)+\frac{7}{16}\ln [\cosh k h(\rt)-c_2\sinh k h(\rt)],\\
h(\rt)&\equiv&\ln\left(\frac{\rt-r_0}{\rt+r_0}\right),\qquad k\equiv\sqrt{4-\frac{7}{16}c_1^2}.
\end{eqnarray*}
In these expressions,  $c_1$, $c_2$ and $r_0$ represent the three parameters of the solution, and the seven-form gauge field given by (\ref{Brax_gauge}) represents that of an electrically charged six-brane; for a magnetic six-brane, the corresponding dilaton field and gauge field can be obtained by performing an electromagnetic duality transformation on the above solution \cite{Brax}.

The parameter $c_1$ was argued in \cite{Brax} to be related to the vacuum expectation value of the open string tachyon stretching between the $D6$- and anti-${D6}$-branes.\footnote{However, the physical significance of this parameter in four dimensions is still unclear.} To establish the correspondence with our solution, we will set this parameter to zero. In addition, the other parameters are taken to be
\bd
c_2=\frac{m}{\sqrt{m^2-l^2}} \an r_0=\frac{\sqrt{m^2-l^2}}{2}.
\ed
Defining a new radial coordinate $r$ by
\bd
\rt=\half(r-m+\sqrt{r^2-2mr+l^2}),
\ed
it can then be checked that the magnetic solution obtained by dualizing (\ref{Brax_metric})~--~(\ref{Brax_dilaton}) is equivalent to (\ref{coincident_metric})~--~(\ref{coincident_dilaton}).
Thus, we see that our solution (\ref{D6D6})~--~(\ref{string_dilaton}) contains, as a special case, the coincident $D6$-anti-${D6}$-brane system of \cite{Brax}.

\newsection{Conclusion}

In this paper, we have presented a solution-generating technique
which was first used by Davidson and Gedalin \cite{Davidson} to
generate black dihole solutions carrying equal but opposite
charges. For any stationary, axisymmetric solution to the vacuum
Einstein equations, we can find a corresponding static,
axisymmetric solution to the Einstein-Maxwell-Dilaton equations
via this technique.

As an application of the technique, we have constructed a new
solution starting from the Kerr-NUT solution. A detailed analysis
reveals that for $l\le m$, the solution describes a pair of
extremal dilatonic black holes lying on the symmetry axis. They
carry unbalanced magnetic charges, with the net charge governed by
the NUT-parameter $l$.

There are a few avenues for future research. Chattaraputi et al.~\cite{Chattaraputi} have recently found
oppositely charged dihole solutions in U(1)$^4$ gauge
theory---a generalization of Einstein-Maxwell-Dilaton theory
consisting of four abelian gauge fields and three scalar fields.
When embedded in string or M-theory, these solutions describe a
variety of intersecting brane-antibrane configurations.
It would be worth finding the corresponding solutions with
unbalanced electric and/or magnetic charges.

It would also be of interest to find dihole solutions describing
non-extremal black holes, as well as diholes in de Sitter and anti-de
Sitter space. Another challenging problem is the construction of
 diholes in  higher-dimensional Einstein-Maxwell-Dilaton theory.
When embedded in string theory, these solutions would describe $Dp$-anti-$Dp$-brane configurations for
$p\leq5$.

\bigbreak\bigskip\bigskip\centerline{{\bf Acknowledgement}}
\nobreak\noindent We thank Roberto Emparan for constructive
criticisms on an earlier version of the manuscript.

\bigskip

{\renewcommand{\Large}{\normalsize}
}
\end{document}